%% file: constraints_letter_revised.tex
\documentclass[aps,prd,twocolumn,preprintnumbers,footinbib,amsmath,amssymb]{revtex4-2}

\usepackage{graphicx}
\usepackage[pdftex, pdftitle={Constraints on the Dirac spectrum from
  chiral symmetry restoration},hidelinks]{hyperref}

\newcommand{\f}[2]{\frac{#1}{#2}}
\newcommand{\tf}[2]{{\textstyle\f{#1}{#2}}}

\newcommand{\de}{\partial}

\newcommand{\la}{\langle}
\newcommand{\ra}{\rangle}

\newcommand{\bpsi}{\bar{\psi}}

\newcommand{\susc}{\mathcal{A}}
\newcommand{\coeff}{\mathcal{C}}
\newcommand{\ww}{\bar{\mathcal{W}}}
\newcommand{\wwh}{\hat{\mathcal{W}}}
\newcommand{\ff}{f}
\newcommand{\fft}{\tilde{f}}
\newcommand{\ffh}{\hat{f}}
\newcommand{\hh}{h}
\newcommand{\topoa}{\mathcal{T}_1}
\newcommand{\topob}{\mathcal{T}_2}
\newcommand{\topoc}{\mathcal{T}_3}
\newcommand{\Iu}{I^{(1)}}
\newcommand{\Id}{I^{(2)}}
\newcommand{\Iq}{I_{Q^2}^{(1)}}
\newcommand{\In}{I_{N_0}^{(1)}}

\newcommand{\chit}{\chi_t}
\newcommand{\lvol}{\mathrm{V}_4}
\newcommand{\svol}{\mathrm{V}_3}

\newcommand{\sous}{V}
\newcommand{\soup}{W}

\newcommand{\ssi}{j_S}
\newcommand{\seta}{j_P}
\newcommand{\ssieta}{j_{S,P}}

\newcommand{\ssit}{J_S}
\newcommand{\setat}{J_P}

\newcommand{\spi}{\vec{\jmath}_P}
\newcommand{\sde}{\vec{\jmath}_S}
\newcommand{\spia}{j_{P\alpha}}
\newcommand{\sdea}{j_{S\alpha}}

\newcommand{\des}{K}
\newcommand{\U}{\mathcal{U}}
\newcommand{\fdet}{F}
\newcommand{\derW}{\omega}

\begin{document}

\raggedbottom

\title{Constraints on the Dirac spectrum from chiral symmetry
  restoration}

\author{Matteo Giordano} \email{giordano@bodri.elte.hu}
\affiliation{Institute of Physics and Astronomy, ELTE E\"otv\"os
  Lor\'and University, P\'azm\'any P\'eter s\'et\'any 1/A, H-1117,
  Budapest, Hungary}

\date{\today}

\begin{abstract}
  I derive constraints on the Dirac spectrum in the chirally symmetric
  phase of a gauge theory with two massless fermion flavors. Using
  only general properties of correlation functions of scalar and
  pseudoscalar bilinears, I prove that in the chiral limit of
  vanishing fermion mass $m$ the corresponding susceptibilities and
  all their derivatives with respect to $m^2$ must be finite.  I then
  use the resulting spectral constraints to show that effective
  breaking of the anomalous $\mathrm{U}(1)_A$ symmetry is allowed in
  the $\mathrm{SU}(2)_A$ symmetric phase in the chiral limit, and
  leads to distinctive spectral features: (i) the spectral density
  must develop a singular $O(m^4)/\lambda$ peak as $m\to 0$, (ii) the
  two-point eigenvalue correlator of near-zero modes must be singular,
  and (iii) near-zero modes cannot be localized. Moreover, in the
  symmetric phase the topological charge distribution must be
  indistinguishable from that of an ideal gas of instantons and
  anti-instantons of vanishing density, to leading order in $m$.
\end{abstract}

\maketitle

The presence of two light fermions (the $u$ and $d$ quarks) among the
strongly interacting fundamental constituents of matter has deep
consequences, best understood theoretically by studying Quantum
Chromodynamics (QCD) in the chiral limit where $u$ and $d$ are exactly
massless.  In this limit QCD has an exact
$\mathrm{SU}(2)_L\times \mathrm{SU}(2)_R$ chiral symmetry,
spontaneously broken to its diagonal $\mathrm{SU}(2)_V$ part at low
temperature.  At higher temperature hadronic matter turns into a
plasma of quarks and gluons, and the broken $\mathrm{SU}(2)_A$ part of
the symmetry gets restored. However, the nature of the
finite-temperature transition and the properties of the
high-temperature symmetric phase are still open questions. A
particularly contentious point is the fate of the anomalous
$\mathrm{U}(1)_A$ symmetry and the related role of gauge-field
topology. These affect the nature of the
transition~\cite{Pisarski:1983ms,Pelissetto:2013hqa}; due to the
vicinity of the chiral and physical points, they can also have
phenomenological consequences in collider
experiments~\cite{Shuryak:1993ee,Kapusta:1995ww,Huang:1995fc,
  Bass:2018xmz} and at the cosmological level~\cite{Bonati:2015vqz,
  Petreczky:2016vrs,Borsanyi:2016ksw,Athenodorou:2022aay}.

There are arguments both in favor and against the necessity of
effective $\mathrm{U}(1)_A$ restoration as a consequence of
$\mathrm{SU}(2)_A$ restoration. Arguments in favor are based on the
consequences of $\mathrm{SU}(2)_A$ restoration for the Dirac spectrum,
and on how these reflect on the $\mathrm{U}(1)_A$ order
parameters. Under plausible analyticity assumptions on the mass
dependence of observables in the symmetric phase, a smooth near-zero
behavior of the spectral density of Dirac modes implies the absence of
$\mathrm{U}(1)_A$-breaking effects in scalar and pseudoscalar
susceptibilities~\cite{Cohen:1996ng,Aoki:2012yj,Kanazawa:2015xna}.
Arguments against effective $\mathrm{U}(1)_A$ restoration are instead
based on assuming commutativity of the chiral and thermodynamic
limits, which generally leads one to expect $\mathrm{U}(1)_A$-breaking
effects from the topology-related zero modes~\cite{Evans:1996wf,
  Lee:1996zy,Kanazawa:2014cua,Kanazawa:2015xna}. This requires,
however, the presence in the spectral density of an implausible
singular delta at the origin at finite quark
mass~\cite{Azcoiti:2023xvu}, calling into question the full
restoration of $\mathrm{SU}(2)_A$ symmetry if $\mathrm{U}(1)_A$
remains broken~\cite{Azcoiti:2021gst}.

This issue could be settled in principle by numerical investigations
on the lattice, but numerical results are also contradictory, with
some supporting~\cite{Tomiya:2016jwr,Brandt:2016daq,Aoki:2020noz}, and
some disfavoring~\cite{HotQCD:2012vvd,Buchoff:2013nra} the effective
restoration of $\mathrm{U}(1)_A$. Moreover, recent investigations (see
also~\cite{Edwards:1999zm}) point at the presence of a singular
near-zero peak in the spectral
density~\cite{Cossu:2013uua,Dick:2015twa,Alexandru:2015fxa,Alexandru:2019gdm,
  Alexandru:2021pap,Alexandru:2021xoi,Meng:2023nxf,
  Tomiya:2016jwr,Aoki:2020noz, Kovacs:2017uiz,
  Vig:2021oyt,Ding:2020xlj,Kaczmarek:2021ser,
  Kaczmarek:2023bxb,Alexandru:2024tel}, suggesting that neither the
smoothness assumption used in
Refs.~\cite{Cohen:1996ng,Aoki:2012yj,Kanazawa:2015xna,Azcoiti:2023xvu}
nor the commutativity assumption used in Refs.~\cite{Evans:1996wf,
  Lee:1996zy,Kanazawa:2014cua,Kanazawa:2015xna,Azcoiti:2023xvu} may be
correct. There is currently no agreement on the physical meaning of
the peak and on its fate in the chiral limit. Some ascribe it to a new
phase of QCD~\cite{Alexandru:2015fxa,Alexandru:2019gdm,
  Alexandru:2021pap,Alexandru:2021xoi,Meng:2023nxf} while others
suggest a topological origin~\cite{Edwards:1999zm,
  Dick:2015twa,Kovacs:2017uiz,Ding:2020xlj,Vig:2021oyt,
  Kaczmarek:2021ser,Kaczmarek:2023bxb}; some expect it to become a
Dirac delta at zero in the chiral
limit~\cite{Dick:2015twa,Ding:2020xlj}, while others claim that it
disappears already at finite quark
mass~\cite{Tomiya:2016jwr,Aoki:2020noz}. In this context, a recent
calculation in a QCD-inspired random matrix model for topology-related
zero and near-zero modes supports $\mathrm{SU}(2)_A$ restoration with
$\mathrm{U}(1)_A$ remaining broken due to a singular spectral
peak~\cite{Kovacs:2023vzi}. While the validity of results obtained in
a mixed-action
setup~\cite{Dick:2015twa,Alexandru:2015fxa,Alexandru:2019gdm,
  Meng:2023nxf,Kaczmarek:2021ser}, using a nonchiral action in the
sea sector and a chiral one in the valence sector, has been questioned
due to the presence of large lattice
artifacts~\cite{Tomiya:2016jwr,Aoki:2020noz},
Refs.~\cite{Ding:2020xlj,Alexandru:2024tel} show how the peak emerges
toward the continuum limit even using the same nonchiral action in
both sectors. While waiting for more precise numerical results to
definitively answer these questions, an analytic first-principles
study of the possible consequences of the spectral peak would be
useful.

Several studies of QCD toward the chiral limit have appeared in the
literature in recent years, with particular attention to spectral
aspects~\cite{Azcoiti:2023xvu,Ding:2020xlj,Kaczmarek:2021ser,
  Carabba:2021xmc,Ding:2023oxy,Kovacs:2023vzi,
  giordano_GT_lett,Giordano:2022ghy,Kaczmarek:2023bxb,
  Alexandru:2024tel}. The constant advancement in computational power
may soon allow for a direct numerical study of the chiral limit,
including with chiral discretizations of the Dirac
operator~\cite{Ginsparg:1981bj,Kaplan:1992bt,Neuberger:1997fp,
  Hasenfratz:1998jp,Luscher:1998pqa}. A reassessment of the
constraints of $\mathrm{SU}(2)_A$ and $\mathrm{U}(1)_A$ restoration on
the Dirac spectrum, separating the consequences of the assumption of
symmetry restoration from those of extra, technical assumptions, is
therefore timely.

Restoration of chiral symmetry is signalled by correlation functions
of local operators related by chiral transformations becoming equal in
the chiral limit. (Here and below, this is always understood to be
taken after the thermodynamic limit.)  Susceptibilities, i.e.,
connected correlation functions integrated over spacetime (normalized
by the four-volume), are then also equal in the chiral limit, provided
that the zero-momentum limit corresponding to spacetime integration
commutes with the chiral limit, which is expected to be the case in
the high-temperature symmetric phase where the correlation length of
the system is finite.  (The critical temperature is excluded in the
case of continuous transitions, since the correlation length diverges
there.)  In the scalar and pseudoscalar sector, susceptibilities can
be expressed in terms of the Dirac eigenvalues only, leading to
necessary conditions required of the Dirac spectrum for
$\mathrm{SU}(2)_A$ restoration. I now outline their derivation;
detailed calculations are reported in the Supplemental
Material~\footnote{See Supplemental Material below for details on
  technical assumptions and proofs of the main statements.}.

I consider finite-temperature $\mathrm{SU}(N_c)$ gauge theories with
two light fermions of mass $m$, and possibly further massive fermions,
regularized on a lattice. Lattice units are used throughout this
paper. I employ a discretized massless Dirac operator $D$ obeying the
Ginsparg-Wilson (GW) relation $\{D,\gamma_5\} = 2DR\gamma_5 D$ with
$R$ local~\cite{Ginsparg:1981bj}. This operator has an exact lattice
chiral symmetry~\cite{Luscher:1998pqa} that reduces to the usual one
in the continuum limit. Denoting by $\psi$ and $\bar{\psi}$ two sets
of Grassmann variables, carrying spacetime, color, and flavor indices,
all suppressed in what follows, the relevant integrated densities are
\begin{equation}
  \label{eq:densities}
\begin{aligned}
  S &= \bpsi (1-DR) \psi \,, &&& P &= \bpsi
  (1-DR) \gamma_5\psi\,, \\
  \vec{P} &= \bpsi (1-DR) \vec{\sigma}\gamma_5\psi \,, &&& \vec{S} &=
  \bpsi (1-DR) \vec{\sigma}\psi \,,
\end{aligned}
\end{equation}
where the Pauli matrices $\vec{\sigma}$ act in flavor space, and
summation over suppressed indices is understood.  Restoration of
chiral symmetry for scalar and pseudoscalar susceptibilities can be
expressed using their generating function
$\mathcal{W}(\sous,\soup;m) = \lim_{\lvol\to\infty}\f{1}{\lvol} \ln
\mathcal{Z}(\sous,\soup;m)$, where $\lvol=\svol/T$ is the lattice
four-volume, $\svol$ its spatial volume and $1/T$ its temporal
extension, equal to the inverse temperature, and
$\mathcal{Z}(\sous,\soup;m)$ is the partition function in the presence
of sources $\sous=(\ssi,\spi)$ and $\soup=(\seta,\sde)$,
\begin{equation}
  \label{eq:partfunc}
  \begin{aligned}
    \mathcal{Z}(\sous,\soup;m) &= \int DU\, e^{-S(U)}
    \fdet (U;\sous,\soup;m)\,,\\
    \fdet (U;\sous,\soup;m) &= \int D\psi D\bar{\psi} \,e^{-
      \bar{\psi} D_m (U) \psi - \des( \psi,\bar{\psi},U; \sous,\soup)
    },\\
    \des( \psi,\bar{\psi},U; \sous,\soup) &= \ssi S + i\spi\cdot
    \vec{P} + i \seta P - \sde\cdot \vec{S}\,,
  \end{aligned}
\end{equation}
with $D_m = D + m\left(1-DR\right)$. The dependence of $\mathcal{W}$
on $T$, and of $\mathcal{Z}$ on $T$ and $\svol$, is understood. Here
$U$ and $DU$ denote collectively the gauge links and the corresponding
Haar integration measure, and $e^{-S}$ includes the contributions of
the discretized Yang-Mills action and of the massive-fermion
determinants.  Usual (anti)periodic boundary conditions are
understood. Expectation values of gauge-field functionals are defined
as
$\la \mathcal{O}\ra = \int DU e^{-S} \fdet|_0\,
\mathcal{O}/\mathcal{Z}|_0$, where $|_0$ denotes setting sources to
zero; their connected part is denoted $\la\cdot\ra_c$.
Susceptibilities are the derivatives of $\mathcal{W}$ at zero sources.
For present purposes, $\mathcal{W}$ (in its various guises discussed
below) is a formal power series that can always be truncated to an
arbitrarily large but finite order, and so it can be treated in
practice as an analytic function of the sources; in particular,
derivation with respect to the sources can be freely exchanged with
the chiral limit.

The multiplets $(S,i\vec{P})$ and $(iP,-\vec{S})$ are irreducible
under chiral transformations $\U$, whose effect on $\des$ is simply
$\des(\psi,\bar{\psi},U;\sous,\soup)\to
\des(\psi,\bar{\psi},U;R_{\U}\sous,R_{\U}\soup)$, with suitable
$R_{\U}\in \mathrm{SO}(4)$. Thanks to $\mathrm{SU}(2)_V$ symmetry,
which is exact also at $m\neq 0$, $\mathcal{W}$ depends on $\spi$ and
$\sde$ only through $\spi^{\,2}$, $\sde^{\,2}$, and $\spi\cdot \sde$,
or equivalently $Y= (Y_1,Y_2,Y_3)=(\sous^2,\soup^2,\sous\cdot \soup)$,
\begin{equation}
  \label{eq:Wvar1}
  \mathcal{W}(\sous,\soup;m) = 
  \ww(\ssi,\seta;Y;m)\,.  
\end{equation}
Notice that $\mathrm{SU}(2)_V$ cannot break spontaneously for GW
fermions at any nonzero $m$~\cite{Giordano:2023spj}. The axial part
$\mathrm{SU}(2)_A$ is instead explicitly broken at nonzero $m$.

Full symmetry restoration in the chiral limit for the
quantities under investigation requires
\begin{equation}
  \label{eq:chrest}
  \lim_{m\to 0} \left[\mathcal{W}(R_{\U}\sous,
  R_{\U}\soup;m)-\mathcal{W}(\sous,\soup;m)\right]=0\,.
\end{equation}
Since Eq.~\eqref{eq:chrest} must hold for arbitrary transformations,
in the chiral limit $\mathcal{W}$ must not depend explicitly on
$\ssieta$,
\begin{equation}
  \label{eq:zchi}
  \lim_{m\to 0} \de_{\ssieta}\ww(\ssi,\seta;Y;m) = 0\,,
\end{equation}
where $\de_x = \de/\de x$. As $\mathcal{W}(\sous,\soup;m)$ depends on
$\ssi$ and $m$ only through the combination $\ssi+m$, it is identical
to
$\mathcal{W}_0(\tilde{\sous},\soup)=\lim_{\lvol\to\infty}
\f{1}{\lvol}\ln\mathcal{Z}(\tilde{\sous},\soup;0)$, where
$\tilde{\sous}=(\ssi+m,\spi)$ is a modified, mass-dependent source.
Since $D$ and the integration measure are invariant under (nonsinglet)
chiral transformations, $\mathcal{W}_0(\tilde{\sous},\soup)$ depends
only on $\mathrm{SO}(4)$ invariant combinations of $\tilde{\sous}$ and
$\soup$,
\begin{equation}
  \label{eq:so4inv}
  \begin{aligned}
    \mathcal{W}(\sous,\soup;m) &= \mathcal{W}_0(\tilde{\sous},\soup) =
    \wwh(\tilde{\sous}^2,\soup^2,2\tilde{\sous}\cdot \soup) \\ &=
    \wwh(m^2 + u,w,\tilde{u}) \,,
  \end{aligned}
\end{equation}
where $u=2m\ssi + \sous^2$, $w=\soup^2$, and
$\tilde{u}=2(m\seta +\sous\cdot \soup)$. (This relation holds exactly
at any finite $m$, even if $\mathrm{SU}(2)_A$ breaks down
spontaneously in the chiral limit, as it follows from the symmetry
properties of the massless partition function in a finite volume in
the presence of nonzero sources.) Then,
$\de_{\ssi}\ww = 2m\de_{\sous^2}\ww$ and
$\de_{\seta}\ww = m\de_{\sous\cdot \soup}\ww$, and chiral symmetry
will be restored if the derivatives of $\ww$ with respect to
$\sous^2$, $\soup^2$, and $\sous\cdot \soup$ at zero sources are
finite in the chiral limit.  The converse is also true: since
\begin{equation}
  \label{eq:so4inv2}
  \begin{aligned}
    & \de_m \ww(\ssi,\seta;Y;m) = \de_m\mathcal{W}(\sous,\soup;m) =
    \de_{\ssi}\mathcal{W}(\sous,\soup;m) \\ & = \left(\de_{\ssi} +
      2\ssi\de_{\sous^2} + \seta\de_{\sous\cdot
        \soup}\right)\ww(\ssi,\seta;Y;m)\,,
  \end{aligned}
\end{equation}
taking further derivatives with respect to $Y_i$ (that commute with
the differential operator appearing on the second line), setting
sources to zero, and taking the chiral limit, Eq.~\eqref{eq:zchi}
implies
\begin{equation}
  \label{eq:dersusc}
  \begin{aligned}
    &\lim_{m\to 0} \de_m \left({\textstyle\prod_i}
      \de_{Y_i}^{n_i}\right)\ww(\ssi,\seta;Y;m)|_0\\
    &= \lim_{m\to 0} \de_{\ssi} \left({\textstyle\prod_i}
      \de_{Y_i}^{n_i}\right)\ww(\ssi,\seta;Y;m)|_0=0\,,
  \end{aligned}
\end{equation}
and so
$\susc_{n_1,n_2,n_3}(m) = (\prod_i\de_{Y_i}^{n_i})
\ww(\ssi,\seta,Y,m)|_{0} $ are finite (meaning nondivergent, here and
below) in the chiral limit.  Taking now repeated derivatives of $\ww$
with respect to $m$ and $Y_i$ and using Eqs.~\eqref{eq:so4inv} and
\eqref{eq:so4inv2} one finds
\begin{equation}
  \label{eq:arg}
  \de_m^n\susc_{n_1,n_2,n_3}(m) = \sum_{a=0}^{\lfloor
    \f{n}{2}\rfloor} C_a^{(n)} (2m)^{n-2a} \susc_{n_1+n-a,n_2,n_3}(m) \,,
\end{equation}
with mass-independent coefficients $C_a^{(n)}$.  By the result above,
odd derivatives vanish in the chiral limit, and even derivatives are
finite. Chiral symmetry restoration requires then all
$\mathcal{A}_{n_1 ,n_2 ,n_3}$ (including the free energy
$\mathcal{A}_{0,0,0}$) to be infinitely differentiable in $m^2$ at
zero (``$m^2$-differentiable'', for short), i.e., to be power series
in $m^2$ (possibly with zero radius of convergence, but this does not
affect the following arguments) up to terms vanishing faster than any
power. This implies that susceptibilities are $m^2$-differentiable, or
$m$ times a $m^2$-differentiable function, depending on whether they
involve an even or odd number of $S$ and $P$ [see
Eq.~\eqref{eq:so4inv}]. This condition, which is commonly assumed, is
here proved to be necessary for $\mathrm{SU}(2)_A$ restoration, using
only general properties of correlators in the symmetric phase.

By a simple generalization of the argument, also connected correlation
functions of integrated scalar and pseudoscalar densities and
(mass-independent) local fields built out of gauge links, and their
spacetime integrals, must be $m^2$-differentiable for chiral symmetry
to be restored. Since gauge fields are unaffected by chiral
transformations, one expects this result to extend to any functional
of the gauge fields; however, for nonlocal functionals (such as the
spectral density of the Dirac operator) this will be treated as an
additional assumption, that I will refer to as ``nonlocal
restoration''.

I now specialize to $\gamma_5$-Hermitian $D$ obeying the GW relation
with $R=\f{1}{2}$~\cite{Kaplan:1992bt,Neuberger:1997fp}, whose
spectrum consists of $N_\pm$ zero modes of definite chirality $\pm 1$,
$N_2$ modes with eigenvalue $2$, and $N$ complex-conjugate pairs of
eigenvalues $\lambda_n^2/2 \pm i\lambda_n \sqrt{1-\lambda_n^2/4}$,
$\lambda_n\in (0,2)$.  I denote by $N_0 = N_++N_-$ and $Q = N_+-N_-$
the total number of zero modes and the topological charge,
respectively, and by $n_0 = \lim_{\lvol\to\infty}{\la N_0\ra}/{\lvol}$
and $\chit = \lim_{\lvol\to\infty}{\la Q^2\ra}/{\lvol}$ the zero-mode
density and the topological susceptibility.  Due to $CP$ symmetry,
$\mathcal{W}$ must be invariant under $\seta,\spi\to -\seta,-\spi$, so
one can express it as a power series in $u$, $w$, and $\tilde{u}^2$,
\begin{equation}
  \label{eq:genfunc}
  \begin{aligned}
    \mathcal{W} &= \coeff_{0} +u\coeff_u +
    w\coeff_{w} + \tilde{u}^2\coeff_{\tilde{u}^2} \\
    &\phantom{=}+ \tf{1}{2}\left(u^2\coeff_{uu} + 2 u w\coeff_{u w} +
      w^2\coeff_{w w} \right. \\ &\phantom{=} \left.+
      2u\tilde{u}^2\coeff_{u\tilde{u}^2}+ 2
      w\tilde{u}^2\coeff_{w\tilde{u}^2} +
      \tilde{u}^4\coeff_{\tilde{u}^2\tilde{u}^2}\right) + \ldots\,.
  \end{aligned}
\end{equation}
Chiral symmetry restoration requires all the coefficients to be finite
in the chiral limit. To first order one finds
\begin{equation}
  \label{eq:genfunc2}
  \begin{aligned}
    \coeff_u &= \f{\chi_\pi}{2} = \f{n_0}{m^2} + 2\Iu[\ff] \,,\\
    \coeff_{w} &= \f{\chi_\delta}{2} = -\f{n_0}{m^2} + 2\Iu[\fft] \,,\\
    \coeff_{\tilde{u}^2} &=
    \f{\chi_\pi-\chi_\delta}{8m^2}-\f{\chit}{2m^4} = \f{1}{m^2}\left(
      \f{n_0-\chit}{2m^2}+m^2\Iu[\ff^2] \right) \,,
  \end{aligned}
\end{equation}
where
$\rho(\lambda;m) = \lim_{\lvol\to\infty}\left\la
  \rho_U(\lambda)\right\ra$ is the spectral density,
\begin{equation}
  \label{eq:specdens}
  \begin{aligned}
    \rho_U(\lambda) & =\tf{1}{\lvol}
    {\textstyle\sum_{n=1}^N}\delta(\lambda-\lambda_n)
    \,, \\
    \Iu[g] & =\int_0^2 d\lambda \,
    g(\lambda)\rho(\lambda;m)\,,\\
    \ff(\lambda;m)&= \f{\hh(\lambda)}{\lambda^2 + m^2\hh(\lambda)} \,,
    \qquad \hh(\lambda)= 1-\tf{\lambda^2}{4}\,,
\end{aligned}
\end{equation}
and $\fft = \ff-2m^2\ff^2$. The dependence of $\rho$ and other
spectral quantities on $T$ is understood. Since both terms in
$\coeff_u$ are positive, they must be separately finite,
\begin{equation}
  \label{eq:const1}
  \lim_{m\to 0} \f{n_0}{m^2}<\infty\,, \qquad
  \lim_{m\to 0} 
  \Iu[\ff]<\infty\,.
\end{equation}
This automatically implies the vanishing of the chiral condensate
$\Sigma = - m\chi_\pi$ in the chiral limit. Since $m^2\ff^2 \le \ff$
for $\lambda\in [0,2]$, $\chi_\delta$ is automatically finite if
$\chi_\pi$ is. Finiteness of $\coeff_{\tilde{u}^2}$ requires that the
term in brackets be $O(m^2)$, and so that the
$\mathrm{U}(1)_A$-breaking parameter
$\Delta = \lim_{m\to 0}\f{\chi_\pi-\chi_\delta}{4}$ equal
\begin{equation}
  \label{eq:const1chi}
  \Delta = \lim_{m\to 0} \left( \f{n_0}{m^2} + 2m^2\Iu[\ff^2] \right)=\lim_{m\to
    0}\f{\chit}{m^2} \,,
\end{equation}
and $\chit=O(m^2)$. Without further assumptions on $n_0$ and $\rho$,
$\mathrm{SU}(2)_A$ restoration is compatible with $\mathrm{U}(1)_A$
breaking.

It is argued, and well supported by numerical results, that
$\la N_+ N_-\ra= 0$~\cite{Blum:2001qg}, implying
$\la N_0^2\ra=\la Q^2\ra$ and $n_0=0$; $\Delta$ is then entirely
determined by the density of near-zero modes, since the region
$\lambda>\delta$ does not contribute to $m^2\Iu$ in the chiral limit
for arbitrary $\delta>0$. If $\rho(\lambda;0)$ is well defined, and
$\rho(\lambda;m)$ admits an expansion in positive powers of $\lambda$,
$\rho(\lambda;m)=\sum_{n=0}^\infty \rho_n(m)\lambda^n$, convergent
within some mass-independent radius (as assumed in
Refs.~\cite{Aoki:2012yj,Kanazawa:2015xna}), then the
$\mathrm{SU}(2)_A$ restoration condition, Eq.~\eqref{eq:const1}, and
positivity of $\rho$ require only $\rho_0(m) = O(m)$ and
$\rho_1(m) = O(1/\ln |m|)$. Effective $\mathrm{U}(1)_A$ restoration,
i.e., $\Delta=0$, requires $\rho_0(m)=o(m)$: this is the case if
$\rho_n$ are power series in $m^2$, as assumed in
Refs.~\cite{Aoki:2012yj,Kanazawa:2015xna}, and follows if restoration
is nonlocal.  A power-series behavior of $\rho$ as a function of both
$\lambda$ and $m^2$ leads then to $\mathrm{U}(1)_A$
restoration~\cite{Aoki:2012yj,Kanazawa:2015xna}.

However, such a behavior is called into question by recent results
showing a singular near-zero peak in
$\rho$~\cite{Edwards:1999zm,Cossu:2013uua,Dick:2015twa,
  Alexandru:2015fxa,Alexandru:2019gdm,Alexandru:2021pap,
  Alexandru:2021xoi,Meng:2023nxf,Tomiya:2016jwr,Aoki:2020noz,
  Kovacs:2017uiz,Ding:2020xlj,Vig:2021oyt,Kaczmarek:2021ser,
  Kaczmarek:2023bxb,Alexandru:2024tel}. For a spectral density
dominated by a power-law behavior $\rho\simeq C(m)\lambda^{\alpha(m)}$
at small $\lambda$, with $|\alpha(m)|<1$ at small nonzero $m$ and
$\alpha(0)\neq 1$, Eq.~\eqref{eq:const1} requires that
$C(m) = \f{\cos \left({\alpha(m)}\pi /{2} \right)}{
  (1-\alpha(0))\pi/2}m^{1-\alpha(0)} (c+o(1))$, resulting in
$\Delta=c$, and so in effective $\mathrm{U}(1)_A$ breaking if
$c\neq 0$. In this case $\rho$ cannot be $m^2$-differentiable, unless
$\alpha(0)=-1$ with $\alpha(m)$ and $C(m)/m^2$ $m^2$-differentiable.
This leads to a surprising and highly nontrivial result: effective
$\mathrm{U}(1)_A$ breaking in the symmetric phase requires a singular
peak in $\rho$ tending to $C/\lambda$ with $C=O(m^4)$ in the chiral
limit, if restoration is nonlocal and $\rho$ an ordinary function. (In
Ref.~\cite{Aoki:2012yj} a nonanalytic behavior
$\rho \sim \lambda^\alpha$ was considered, but only for
mass-independent $\alpha > 0$. A behavior $\rho\sim 1/\lambda$ was
proposed in
Refs.~\cite{Alexandru:2019gdm,Alexandru:2021pap,Alexandru:2021xoi} in
a different context.) This is possible only if the chiral and
thermodynamic limits do not commute (otherwise one needs a term
$m^2\Delta \delta(\lambda)$ in $\rho$)~\cite{Azcoiti:2023xvu}.  In
this case one also finds
$2\int_0^2 d\lambda\, C(m)\lambda^{\alpha(m)} = \chit +o(m^2)$ for the
density of peak modes. This supports the topological origin of the
peak~\cite{Edwards:1999zm,
  Dick:2015twa,Kovacs:2017uiz,Ding:2020xlj,Vig:2021oyt,
  Kaczmarek:2021ser,Kaczmarek:2023bxb}, and is consistent with the
instanton model of Ref.~\cite{Kovacs:2023vzi}. Moreover, it shows how
strongly suppressed the peak is in the chiral limit, requiring very
large volumes of order $\chit^{-1} \propto m^{-2}$ to become fully
visible. The disappearance of the peak at finite $m$ reported in
Refs.~\cite{Tomiya:2016jwr,Aoki:2020noz} could then simply be a
finite-volume artifact.

Further constraints are obtained requiring finiteness of the
second-order terms in Eq.~\eqref{eq:genfunc}. The first two are
\begin{equation}
  \label{eq:const2_1}
  \begin{aligned}
    \f{b_{N_0^2}-\chit}{2m^2} - m^2\f{\de}{\de
      m^2}\f{n_0}{m^2}-2m^2\Id[\ff,\ff] &= O(m^2)
    \,,\\
    \Id[\ffh,\ffh]& = O(m^0) \,,
  \end{aligned}
\end{equation}
where $\ffh = \ff-m^2\ff^2$,
$b_{X} = \lim_{\lvol\to\infty}{\la X\ra_c}/{\lvol}$,
\begin{equation}
  \label{eq:const2bis_1}
  \Id[g_1,g_2] 
  =\int_0^2 d\lambda\int_0^2 d\lambda' \,
  g_1(\lambda) g_2(\lambda')\rho_c^{(2)}(\lambda,\lambda';m)\,,
\end{equation}
and $\rho_c^{(2)}$ is the connected two-point function,
\begin{equation}
  \label{eq:const2bis_2}
  \begin{aligned}
    \rho_c^{(2)}(\lambda,\lambda';m)& =\lim_{\lvol\to\infty}
    \left(\lvol\left\la
        \rho_U(\lambda)\rho_U(\lambda')\right\ra_c\right.\\ &
    \phantom{=()\lim_{\lvol\to\infty} } \left.
      -\delta(\lambda-\lambda')\left\la \rho_U(\lambda)\right\ra
    \right)\,.
  \end{aligned}
\end{equation}
Together with the assumption $\la N_+N_-\ra =0$, these constraints
lead to remarkable results. I assume here that $\rho$ and
$\rho_c^{(2)}$ are ordinary functions; if deltalike singularities are
present, the arguments below apply to their ordinary parts.  Under
general integrability conditions the region
$\lambda,\lambda'>\delta>0$ does not contribute to $m^2\Id$ in the
chiral limit, for arbitrarily small $\delta$. If one assumes
furthermore that
$\rho_c^{(2)}(\lambda,\lambda';m)=A(m)+B(\lambda,\lambda';m)$, with
$B$ vanishing at the origin and obeying the loose bound
$|B(\lambda,\lambda';m)|\le b (\lambda^2 +
\lambda^{\prime\,2})^{\f{\beta}{2}}$ with $0<\beta<1$, the first
constraint implies
\begin{equation}
  \label{eq:BClike}
  -  \pi^2 A(0)  = \lim_{m\to 0}  \f{\chit-b_{N_0^2}}{m^2}
  = \lim_{m\to 0}\lim_{\lvol\to \infty}\f{\la N_0\ra^2}{\lvol m^2} =\Delta'
  \,.
\end{equation}
(This corrects a result of Ref.~\cite{Kanazawa:2015xna} where exact
zero modes were not fully accounted for.) On the other hand, the
second constraint implies that
$(\pi/4)^2 A(0)=\lim_{m\to 0} m^2\Id[\ffh,\ffh] = 0$, so $\Delta'$
vanishes. This is possible only if the measure of the positive random
variable $\f{N_0}{\sqrt{\lvol} |m|}$ is concentrated in zero in the
thermodynamic and chiral limit, implying in turn $\f{\chit}{m^2}\to 0$
and $\Delta=0$. $\mathrm{U}(1)_A$ breaking by $\Delta\neq 0$ in the
$\mathrm{SU}(2)_A$-restored phase requires then a singular behavior at
the origin of the two-point function $\rho_c^{(2)}$ for $m\neq 0$.

The current wisdom is that in the high-temperature phase of QCD, low
modes are localized below a ``mobility edge'', $\lambda_c$, and obey
Poisson statistics~\cite{Giordano:2021qav}. For a purely Poisson
spectrum of $\mathrm{N}_{\mathrm{P}}\lvol$ eigenvalues one has
$\rho_{\mathrm{P}\,c}^{(2)}(\lambda,\lambda') =
-\rho_{\mathrm{P}}(\lambda)\rho_{\mathrm{P}}(\lambda')/
\mathrm{N}_{\mathrm{P}}$~\cite{Guhr:1997ve,Kanazawa:2015xna}, so one
expects
$|\rho_c^{(2)}(\lambda,\lambda';m)| \le C
\rho(\lambda;m)\rho(\lambda';m)$ for some constant $C$ if
$\lambda,\lambda'<\lambda_c$. As localized modes fluctuate
independently of each other, one expects their correlations with modes
beyond the mobility edge to be proportional to their density. It is
then reasonable to assume
$|\rho_c^{(2)}(\lambda,\lambda';m)| \le C' \rho(\lambda;m)$ for
$\lambda<\lambda_c<\lambda'$. If $\lambda_c$ remains finite in the
chiral limit, under the assumptions above $m^2\Id[\ff,\ff]$ vanishes
in the chiral limit (and the second constraint in
Eq.~\eqref{eq:const2_1} is satisfied), so $\Delta'=0$. Barring
deltalike contributions, the persistence of
$\mathrm{U}(1)_A$-breaking topological effects in the symmetric phase
requires then that near-zero modes be not localized, either because
$\lambda_c$ vanishes in the chiral limit, or because another mobility
edge is present near zero at nonzero $m$. (A mobility edge exactly at
zero, as proposed in
Refs.~\cite{Alexandru:2021pap,Alexandru:2021xoi,Meng:2023nxf}, is not
sufficient, as the zero-measure lines $\lambda=0$ and $\lambda'=0$ can
be excluded from the integral defining $\Id$.) The second possibility
seems more likely in the presence of a singular spectral peak of
topological origin. In any case, $\mathrm{U}(1)_A$ breaking requires
strong (singular) repulsion of near-zero modes. Numerical evidence for
non-Poissonian repulsion toward the chiral limit was found using
staggered fermions~\cite{Ding:2020xlj}.

Two more constraints obtained at second order are
\begin{equation}
  \label{eq:const2_2}
  \begin{aligned}
    \left| \lim_{m\to 0}\f{\de}{\de m^2}\f{\chit}{m^2}\right| &<
    \infty\,, &&& \left| \lim_{m\to 0} \f{ \de
        \coeff_{\tilde{u}^2}}{\de m^2} \right| &<\infty \,.
  \end{aligned}
\end{equation}
The first one requires that in the symmetric phase
$\chit = \Delta m^2 + c_1 m^4 + o(m^4)$. This follows (to all orders
in $m^2$) from the argument under Eq.~\eqref{eq:arg}, since $Q^2$
admits a representation as the integral of a local
correlator~\cite{Hasenfratz:1998ri,Luscher:2004fu,Vicari:2008jw}. The
second constraint implies that the term in brackets in
Eq.~\eqref{eq:genfunc2} is of the form $d_1m^2 + d_2m^4 + o(m^4)$;
this also follows to all orders from the argument under
Eq.~\eqref{eq:arg}.

The last two constraints at second order are
  \begin{equation}
  \label{eq:const2_3}
  \lim_{m\to 0} \f{8}{m^2} \int_0^2 d\lambda
  \,\ffh(\lambda;m)\rho_{Q^2\,c}(\lambda;m) 
  = \lim_{m\to 0}
  \left(    \coeff_{uu}-\coeff_{ww}\right) \,,
\end{equation}
where
$\rho_{Q^2\,c}(\lambda;m) = \lim_{\lvol\to\infty}\left\la Q^2
  \rho_U(\lambda)\right\ra_c $, and
\begin{equation}
  \label{eq:const2_4}
  \f{b_{Q^4}-\chit}{m^2} = O(m^2) \,.
\end{equation}
No constraint on $\rho$ is found at second order; no more can be
obtained, as higher-order coefficients involve only higher-point
eigenvalue correlators. In Eq.~\eqref{eq:const2_3},
$\coeff_{uu}-\coeff_{ww}\propto \la (iP_a)^2 (iP_b)^2\ra - \la S_a^2
S_b^2\ra$ ($a\neq b$), so if this is nonzero one has
$\mathrm{U}(1)_A$-breaking effects originating in non-negligible
correlations between zero and nonzero modes.

The constraint Eq.~\eqref{eq:const2_4} requires that in the chiral
limit, to leading order in $m$ and to the lowest nontrivial cumulant,
the distribution of $Q$ be indistinguishable from that of an ideal
(noninteracting, Poisson-distributed) gas of instantons and
anti-instantons, with equal and vanishingly small densities
$\chit/2=O(m^2)$. In the symmetric phase, the coefficient $b_2$ in the
partition function at finite $\theta$ angle~\cite{Bonati:2013tt} must
then tend to $-1/12$ in the chiral limit.  This leads one to expect an
early onset (near the transition temperature $T_c$) of an ideal
instanton gaslike behavior for physical quark masses (though not
necessarily that of the usual semiclassical dilute instanton
gas~\cite{Gross:1980br,Boccaletti:2020mxu}), as seen in the pure gauge
case~\cite{Bonati:2013tt}. On the other hand, the results of
Ref.~\cite{Bonati:2015vqz} for $b_2$ show large deviations from an
ideal gas up to $2T_c$, which calls for a deeper scrutiny of the size
of finite-$m$ corrections. As will be shown elsewhere, chiral symmetry
restoration actually requires all cumulants to be those of an ideal
gas, a result already obtained in Ref.~\cite{Kanazawa:2014cua}
assuming a free energy analytic in $m^2$.  (This does not mean that
typical gauge configurations must look like an ordinary ideal
instanton gas~\cite{Kanazawa:2014cua}.)

Summarizing, $\mathrm{SU}(2)_A$ restoration does not prevent effective
$\mathrm{U}(1)_A$ breaking in the chiral limit, but puts severe
constraints on the Dirac spectrum for it to happen: the spectral
density must develop a $O(m^4)/\lambda$ peak; the two-point function
of near-zero Dirac modes must be singular; near-zero modes cannot obey
Poisson statistics and be localized; and the distribution of the
topological charge must be the same as that of an ideal instanton gas
with density of order $m^2$. These predictions of distinctive spectral
features, that can lead to a more accurate characterization of the
quark-gluon plasma phase of QCD, should be verified by numerical
investigations.  The necessary and sufficient condition for chiral
symmetry restoration in the scalar and pseudoscalar sector can
probably be generalized to other sectors and to more massless fermion
flavors by suitably adapting the proof.

\mbox{}

\mbox{}

I thank T.~G.~Kov\'acs for discussions and a careful reading of the
manuscript, 
V.~Azcoiti for useful remarks, and C.~Bonanno for discussions. This
work was partially supported by NKFIH grant K-147396.

\bibliographystyle{apsrev4-2}
\bibliography{references_chi_PRD}

\makeatletter
  \close@column@grid
  \clearpage
  \twocolumngrid
\makeatother

\input{constraints_supplemental_revised.tex}

\end{document}

%% file: constraints_supplemental_revised.tex
\setcounter{equation}{0}
\setcounter{figure}{0}
\setcounter{table}{0}

\makeatletter
\renewcommand{\theequation}{S\arabic{equation}}

\begin{widetext}
\begin{center}
\textbf{\large Supplemental Material: Constraints on the Dirac
  spectrum from chiral symmetry restoration} 
\end{center}
\setcounter{page}{1}

\section{Chiral symmetry}
\label{sec:app_gen}

\subsection{Nonsinglet vector and axial transformations}
\label{sec:vatr}

The action of vector, $\mathrm{SU}(2)_V$, and axial,
$\mathrm{SU}(2)_A$, chiral transformations on GW fermions is given by
\begin{equation}
  \label{eq:chitransf}
\begin{aligned}
  \U_V(\vec{\alpha}) :& &&& \psi&\to
  \psi_{\U_V(\vec{\alpha})}=e^{i\vec{\alpha}\cdot\f{\vec{\sigma}}{2}}\psi\,,&&&
  \bpsi&\to \bpsi_{\U_V(\vec{\alpha})}=\bpsi
  e^{-i\vec{\alpha}\cdot\f{\vec{\sigma}}{2}}\,,
  \\
  \U_A(\vec{\alpha}) :& &&& \psi&\to
  \psi_{\U_A(\vec{\alpha})}=e^{i\vec{\alpha}\cdot\f{\vec{\sigma}}{2}\gamma_5}\psi\,,&&&
  \bpsi&\to \bpsi_{\U_A(\vec{\alpha})}=\bpsi
  e^{i\vec{\alpha}\cdot\f{\vec{\sigma}}{2}\hat{\gamma}_5}\,,
\end{aligned}
\end{equation}
respectively, where $\hat{\gamma}_5\equiv (1-2DR)\gamma_5$, with
$\gamma_5$ the usual Dirac matrix, and it is assumed that
$[R,\gamma_5]=0$. Their effect on the multiplets
$O_{\sous}=(S,i\vec{P})$ and $O_{\soup}=(iP,-\vec{S})$ is
$O_{\sous,\soup}(\psi,\bpsi)\to O_{\sous,\soup}(\psi_\U,\bpsi_\U) =
R_{\U}{}^T O_{\sous,\soup}(\psi,\bpsi)$, with
$R_{\U}\in \mathrm{SO}(4)$. Explicitly,
\begin{equation}
  \label{eq:chitransf2}
  \begin{aligned}
    R_{\U_V(\vec{\alpha})} &=
    \begin{pmatrix}
      1 & \vec{0}^{\,T}\\
      \vec{0} & \tilde{R}(\vec{\alpha})
    \end{pmatrix}
    \,,&&& R_{\U_A(\vec{\alpha})} &=
    \begin{pmatrix}
      \cos (|\vec{\alpha}|)& -\sin (|\vec{\alpha}|) \hat{\alpha}^T \\
      \sin (|\vec{\alpha} |) \hat{\alpha} & \Pi_{\hat{\alpha}_\perp} +
      \cos (|\vec{\alpha}|) \Pi_{\hat{\alpha}}
    \end{pmatrix}\,,
  \end{aligned}
\end{equation}
with $\tilde{R}\in\mathrm{SO}(3)$,
\begin{equation}
  \label{eq:chitransf2_bis}
  \tilde{R}(\vec{\alpha})\vec{X} \equiv 
  \Pi_{\hat{\alpha}}\vec{X} + \cos(|\vec{\alpha}|)
  \Pi_{\hat{\alpha}_\perp}\vec{X}
  -\sin(|\vec{\alpha}|)\hat{\alpha}\wedge\vec{X}\,,
\end{equation}
$\vec{\alpha}=|\vec{\alpha}|\hat{\alpha}$ a three-dimensional vector,
$\Pi_{\hat{\alpha}}=\hat{\alpha}\hat{\alpha}^T$, and
$ \Pi_{\hat{\alpha}_\perp}= 1-\Pi_{\hat{\alpha}}$.  Since replacing
$|\vec{\alpha}|\to-|\vec{\alpha}|$ is equivalent to
$\hat{\alpha}\to -\hat{\alpha}$, one can replace $|\vec{\alpha}|\ge 0$
with $\alpha\in\mathbb{R}$ in the expressions above.  The resulting
effect on the source-dependent part of the integrand in
Eq.~\eqref{eq:partfunc},
$ \des(\psi,\bpsi;\sous,\soup)= \sous\cdot O_\sous(\psi,\bpsi) +
\soup\cdot O_\soup(\psi,\bpsi) $, is
\begin{equation}
  \label{eq:chitransf_source}
  \des(\psi,\bpsi;\sous,\soup)\to
  \des(\psi_\U,\bar{\psi}_\U;\sous,\soup)
  =
  (R_{\U} \sous)\cdot O_\sous(\psi,\bpsi) +
  (R_{\U} \soup)\cdot O_\soup(\psi,\bpsi)
  =\des(\psi,\bar{\psi};R_{\U}\sous,R_{\U}\soup)\,.
\end{equation}

\subsection{Chiral symmetry restoration}
\label{sec:chrest}

I derive here the restoration condition, Eq.~\eqref{eq:zchi}.  For an
$\mathrm{SU}(2)_A$ transformation, the request of chiral restoration
in the chiral limit, Eq.~\eqref{eq:chrest}, reads explicitly
\begin{equation}
  \label{eq:act31}
  \begin{aligned}
    & \lim_{m\to 0} \ww(\ssi,\seta;Y ;m) = \lim_{m\to 0}\ww(
    \ssit(\alpha), \setat(\alpha); Y ;m)\,,
  \end{aligned}
\end{equation}
where
\begin{equation}
  \label{eq:chiZ_notation}
  \begin{aligned}
    \ssit(\alpha)&\equiv\cos(\alpha) \ssi-\sin(\alpha) \spia\,, &&&
    \setat(\alpha)&\equiv\cos(\alpha) \seta+\sin(\alpha) \sdea\,,
  \end{aligned}
\end{equation}
with $\spia= \vec{\alpha}\cdot\spi$ and
$\sdea= \vec{\alpha}\cdot\sde$. Since the left-hand side does not
depend explicitly on $\spia$ and $\sdea$, taking the derivative with
respect to $ \spia$ at constant $\sous^2,\sous\cdot \soup$, and with
respect to $ \sdea$ at constant $\soup^2,\sous\cdot \soup$, one finds
\begin{equation}
  \label{eq:dePa}
  \begin{aligned}
    0 &= -\lim_{m\to 0}\left( \de_{\spia} \ww(\ssi,\seta;Y
      ;m)\right)_{\sous^2,\sous\cdot \soup} = \sin(\alpha) \lim_{m\to
      0}{\de_{\ssit}}\ww( \ssit(\alpha),\setat(\alpha);Y ;m)
    \,,\\
    0 &=\phantom{-} \lim_{m\to 0}\left( \de_{\sdea} \ww(\ssi,\seta;Y
      ;m)\right)_{\soup^2,\sous\cdot \soup} = \sin(\alpha) \lim_{m\to
      0}{\de_{\setat}}\ww(\ssit(\alpha),\setat(\alpha);Y ;m) \,.
  \end{aligned}
\end{equation}
Taking now one derivative with respect to $\alpha$ and setting
$\alpha=0$, one finds
\begin{equation}
  \label{eq:act33_1}
  \begin{aligned}
    0 & = \left( \cos(\alpha) + \sin(\alpha) \de_\alpha \right)
    \lim_{m\to 0}\de_{\ssit}\ww(\ssit(\alpha),\setat(\alpha);Y ;m)
    \big|_{\alpha =0} = \lim_{m\to 0}\de_{\ssi}\ww(\ssi , \seta;Y
    ;m)\,,\\
    0&=\left(\cos(\alpha) +\sin(\alpha) \de_\alpha \right)
    \lim_{m\to 0}\de_{\setat}\ww(\ssit(\alpha),\setat(\alpha);Y ;m)
    \big|_{\alpha =0}
    = \lim_{m\to 0}\de_{\seta}\ww(\ssi , \seta;Y ;m)\,. 
  \end{aligned}
\end{equation}

\subsection{Mass derivatives}
\label{sec:massder}

I derive here the mass-derivative formula, Eq.~\eqref{eq:arg}. To do
so, one first notices that, since $\de_m$ commutes with $\de_{Y_i}$,
and both commute with
$\de_{\ssi} + 2\ssi\de_{\sous^2} + \seta\de_{\sous\cdot \soup}$,
Eq.~\eqref{eq:so4inv2} implies
\begin{equation}
  \label{eq:arg_app1}
  \begin{aligned}
    \de_m^n \left({\textstyle\prod_i} \de_{Y_i}^{n_i}\right)
    \ww(\ssi,\seta;Y;m)&= \left(\de_{\ssi} + 2\ssi\de_{\sous^2} +
      \seta\de_{\sous\cdot \soup}\right)^n \left({\textstyle\prod_i}
      \de_{Y_i}^{n_i}\right) \ww(\ssi,\seta;Y;m)\,.
  \end{aligned}    
\end{equation}
Denoting
$\derW(\ssi,\seta;Y;m)\equiv\left({\textstyle\prod_i}
  \de_{Y_i}^{n_i}\right) \ww(\ssi,\seta;Y;m)$, one has, setting
sources to zero,
\begin{equation}
  \label{eq:arg_app1_bis}
  \begin{aligned}
    \de_m^n \derW(\ssi,\seta;Y;m)|_0 = \left(\de_{\ssi}+
      2\ssi\de_{\sous^2}\right)^n \derW(\ssi,\seta;Y;m)|_0= \sum_{ab}
    \tilde{C}^{(n)}_{ba} \de_{\ssi}^b \de_{\sous^2}^a \derW(\ssi,\seta;Y;m)|_0
    \,,
  \end{aligned}
\end{equation}
for suitable mass-independent coefficients $\tilde{C}^{(n)}_{ba}$.
Their explicit expression can be obtained by simple combinatorics, but
is not needed here.  By dimensional analysis, only terms with
$2a+b = n$ can appear in the last passage, and so, setting
$C_a^{(n)}=\tilde{C}^{(n)}_{n-2a\,a}$, and using the fact that
$\derW(\ssi,\seta;Y;m)= \de_u^{n_1} \de_w^{n_2}
(2\de_{\tilde{u}})^{n_3} \wwh(m^2+u,w,\tilde{u}) =
\hat{\derW}(m^2+u,w,\tilde{u})$,
\begin{equation}
  \label{eq:arg_app2}
  \begin{aligned}
    \de_m^n \derW(\ssi,\seta;Y;m)|_0 &= \sum_{a=0}^{\lfloor
      \f{n}{2}\rfloor} \tilde{C}^{(n)}_{n-2a\,a} \de_{\ssi}^{n-2a}
    \de_{\sous^2}^a \derW(\ssi,\seta;Y;m)|_0= \sum_{a=0}^{\lfloor
      \f{n}{2}\rfloor} C^{(n)}_{a} (2m\de_{u})^{n-2a}
    \de_{u}^a\hat{\derW}(m^2+u,w,\tilde{u})|_0
    \\
    &= \sum_{a=0}^{\lfloor \f{n}{2}\rfloor} C^{(n)}_{a}
    (2m)^{n-2a}\de_{\sous^2}^{n-a}\derW(\ssi,\seta;Y;m)|_0\,.
  \end{aligned}
\end{equation}
Noting that
$\de_m^n \derW(\ssi,\seta;Y;m)|_0=\de_m^n\susc_{n_1,n_2,n_3}(m)$ and
$\de_{\sous^2}^{n-a}\derW(\ssi,\seta;Y;m)|_0=\susc_{n_1+n-a,n_2,n_3}(m)$,
Eq.~\eqref{eq:arg} follows.

\section{Generating function}
\label{sec:app_genfunc}

\subsection{Determinant in the presence of sources}
\label{sec:detsource}

For a $\gamma_5$-Hermitian $D$ obeying the GW relation with
$R=\f{1}{2}$, the spectrum consists of $N_\pm$ zero modes of definite
chirality $\pm 1$, $N_2$ modes with eigenvalue $2$, and $N$
complex-conjugate pairs of eigenvalues $\mu_n,\mu_n^*$, with
$\mu_n = \f{\lambda_n^2}{2} + i\lambda_n \sqrt{1-\f{\lambda_n^2}{4}}$
and $\lambda_n\in (0,2)$; in total,
$N_+ + N_- + N_2 + 2N = \mathrm{N}\lvol$ with volume-independent
$\mathrm{N}$ (e.g., $\mathrm{N}=4N_c$ for the overlap operator). The
fermion determinant, Eq.~\eqref{eq:partfunc}, reads explicitly
\begin{equation}
  \label{eq:determinant2}
  \begin{aligned}
    \f{\fdet(U;\sous,\soup;m)}{\fdet(U;0,0;m)} &= e^{N_0 \tf{1}{2}\ln
      \left(\left(1 + \tf{u-w}{m^2}\right)^2 +
        \tf{\tilde{u}^2}{m^4}\right) } e^{iQ\arctan\tf{\tilde{u}}{m^2
        +u-w} }
    \prod_{n=1}^N \left(1+X(\lambda_n;m)\right)\,,\\
    \fdet(U;0,0;m) & = m^{2N_0}4^{N_2} \prod_{n=1}^N \left(\lambda_n^2
      + m^2\hh(\lambda_n)\right)^2\,,
  \end{aligned}
\end{equation}
where $N_0\equiv N_++N_-$, $Q\equiv N_+-N_-$, and
\begin{equation}
  \label{eq:determinant3}
  \begin{aligned}
    X(\lambda;m)&\equiv 2\left(\ff(\lambda;m)u+ \fft(\lambda;m)
      w\right) +\ff(\lambda;m)^2 \left(\left(u-w\right)^2+
      \tilde{u}^2\right)
    \,,\\
    \ff(\lambda;m)&\equiv\f{\hh(\lambda)}{\lambda^2 + m^2\hh(\lambda)}
    \,, \qquad \hh(\lambda)\equiv 1-\tf{\lambda^2}{4}\,,\qquad
    \fft(\lambda;m)\equiv\ff(\lambda;m)-2m^2\ff(\lambda;m)^2\,.
  \end{aligned}
\end{equation}
Details of the derivation will be presented elsewhere. Notice the
following inequalities, valid for $\lambda\in [0,2]$,
\begin{equation}
  \label{eq:determinant4}
  \begin{aligned}
    \f{1}{\lambda^2 + m^2}-\f{1}{4}&\le \ff(\lambda;m) \le
    \f{1}{\lambda^2 + m^2}\,, &&& m^2\ff(\lambda;m)^2 & \le
    \ff(\lambda;m)\,, &&& |\fft(\lambda;m)| &\le \ff(\lambda;m)\,.
  \end{aligned}
\end{equation}

\subsection{Expansion of the generating function}
\label{sec:expansion}

The coefficients of the first-order terms in the expansion of
$\mathcal{W}$, Eq.~\eqref{eq:genfunc}, read explicitly
\begin{equation}
  \label{eq:first_order_expl}
  \begin{aligned}
    \coeff_u &= \f{n_0}{m^2} + 2\Iu[\ff] \,, &&& \coeff_{w} &=
    -\f{n_0}{m^2} + 2\Iu[\fft] \,, &&& \coeff_{\tilde{u}^2} &=
    \f{1}{m^2}\left(\f{n_0-\chit}{2m^2}+m^2\Iu[\ff^2]\right) \,,
  \end{aligned}
\end{equation}
while those of the second-order terms are
\begin{equation}
  \label{eq:second_order_expl}
  \begin{aligned}
    \coeff_{uu}&= \topoa + 4\Id[\ff,\ff] + \f{4}{m^2}\In[\ff] \,,
    \\
    \coeff_{uw}&= -\topoa + 2\Id[\ff,\fft] + 2\Id[\fft,\ff] +
    \f{2}{m^2}\In[\fft]-\f{2}{m^2}\In[\ff] \,,
    \\
    \coeff_{ww} &= \topoa + 4\Id[\fft,\fft] - \f{4}{m^2}\In[\fft] \,,
    \\
    \coeff_{u\tilde{u}^2} &= \topob + \Id[\ff,\ff^2] + \Id[\ff^2,\ff]
    +\f{1}{m^2} \In[\ff^2] +\f{1}{m^4}\In[\ff] -\f{1}{m^4}\Iq[\ff]
    \,,\\
    \coeff_{w\tilde{u}^2} &= - \topob + \Id[\fft,\ff^2] +
    \Id[\ff^2,\fft] -\f{1}{m^2} \In[\ff^2] +\f{1}{m^4}\In[\fft]
    -\f{1}{m^4}\Iq[\fft]
    \,,\\
    \coeff_{\tilde{u}^2\tilde{u}^2} &= \topoc + \Id[\ff^2,\ff^2] +
    \f{1}{m^4}\In[\ff^2] - \f{1}{m^4}\Iq[\ff^2] \,,
\end{aligned}
\end{equation}
where
 \begin{equation}
   \label{eq:second_order_expl_bis}
   \begin{aligned}
     \topoa &= \f{1}{m^4} \left(b_{N_0^2}-n_0 +
       2m^4\Iu[\ff^2]\right)\,,
     \\
     \topob &= \f{1}{m^6}\left(- \f{1}{2} b_{N_0 Q^2} + \chit +
       \f{1}{2}b_{N_0^2} - n_0  \right)\,,\\
     \topoc &= \f{1}{m^8}\left(\f{1}{12} b_{Q^4}- \f{1}{2}b_{N_0 Q^2}+
       \f{2}{3} \chit +\f{1}{4} b_{N_0^2} -\f{1}{2}n_0 \right)\,,
  \end{aligned}
\end{equation}
with 
\begin{equation}
  \label{eq:second_order_expl_bis_more}
  \begin{aligned}
    n_0&\equiv \lim_{\lvol\to \infty} \f{ \la N_0\ra}{\lvol}\,, &&&
    \chit&\equiv\lim_{\lvol\to \infty} \f{ \la Q^2\ra}{\lvol}\,, &&&    b_X
    &\equiv\lim_{\lvol\to \infty} \f{ \la X\ra_c}{\lvol}\,,\\ 
    \la N_0^2 \ra_c &= \la N_0^2 \ra -\la N_0\ra^2\,,&&& \la N_0 Q^2
    \ra_c &= \la N_0 Q^2 \ra -\la N_0\ra \la Q^2\ra\,, &&& \la Q^4
    \ra_c &= \la Q^4 \ra -3 \la Q^2\ra^2\,,
  \end{aligned}
\end{equation}
and
\begin{equation}
  \label{eq:cum_sys_3_0_again}
  \begin{aligned}
    \Iu[g]&\equiv \int_0^2 d\lambda\, g(\lambda) \rho(\lambda;m)\,, &&&
    \Id[g_1,g_2]&\equiv \int_0^2 d\lambda_1\int_0^2 d\lambda_2\,
    g_1(\lambda_1) g_2(\lambda_2)
    \rho^{(2)}_{c}(\lambda_1,\lambda_2;m)\,,
    \\
    \In[g]&\equiv \int_0^2 d\lambda\,
    g(\lambda)\rho^{(1)}_{N_0\,c}(\lambda;m) \,,&&& \Iq[g]&\equiv \int_0^2
    d\lambda\, g(\lambda)\rho^{(1)}_{Q^2\,c}(\lambda;m)\,,\\
    \rho(\lambda;m) &\equiv \lim_{\lvol\to\infty}\left\la
      \rho_U(\lambda)\right\ra\,, &&& \rho_U(\lambda) &\equiv
    \tf{1}{\lvol} {\textstyle\sum_{n=1}^N}\delta(\lambda-\lambda_n)\,,
    \\
    \rho_{N_0\,c}(\lambda;m)&\equiv \lim_{\lvol\to\infty}\left\la N_0
      \rho_U(\lambda)\right\ra -\left\la N_0\right\ra
    \left\la\rho_U(\lambda)\right\ra\,, &&&
    \rho_{Q^2\,c}(\lambda;m)&\equiv \lim_{\lvol\to\infty}\left\la Q^2
      \rho_U(\lambda)\right\ra -\left\la Q^2\right\ra
    \left\la\rho_U(\lambda)\right\ra\,,
  \end{aligned}
\end{equation}
and moreover
\begin{equation}
  \label{eq:second_order_expl3}
  \begin{aligned}
    \rho_c^{(2)}(\lambda,\lambda';m)&\equiv \lim_{\lvol\to\infty}
    \lvol \left\{\left\la \rho_U(\lambda)\rho_U(\lambda')\right\ra -
      \left\la
        \rho_U(\lambda)\right\ra\left\la\rho_U(\lambda')\right\ra
    \right\} -\delta(\lambda-\lambda')\left\la
      \rho_U(\lambda)\right\ra\,, \\ &= \lim_{\lvol\to\infty}
    \f{1}{\lvol}\left\{ \left\la
         {\sum_{\substack{n,n'=1\\n\neq n'}}^N}
        \delta(\lambda-\lambda_n)\delta(\lambda'-\lambda_{n'})\right\ra
      -\left\la {\sum_{n=1}^N} \delta(\lambda-\lambda_n)\right\ra
      \left\la {\sum_{n'=1}^N} \delta(\lambda'-\lambda_{n'})\right\ra
    \right\}\,,
  \end{aligned}
\end{equation}
with the obvious property
$ \rho_c^{(2)}(\lambda',\lambda;m)= \rho_c^{(2)}(\lambda,\lambda';m)$.
Taking linear combinations of the quantities in
Eq.~\eqref{eq:second_order_expl}, and using the explicit expression
for the following mass derivatives,
\begin{equation}
  \label{eq:mder}
  \begin{aligned}
    m^2\f{\de}{\de m^2}\f{ n_0}{m^2}&= \f{b_{N_0^2}-n_0}{ m^2} +
    2\In[\ff]\,, \qquad\qquad \f{\de \chit}{\de m^2} =
    \f{b_{N_0 Q^2}}{ m^2}  + 2\Iq[\ff]\,,\\
    \f{\de}{\de m^2} \rho(\lambda;m) &=
    \f{1}{m^2}\rho_{N_0\,c}(\lambda;m) + 2
    \ff(\lambda;m)\rho(\lambda;m) + 2 \int_0^2d\lambda'\,
    \ff(\lambda';m)\rho^{(2)}_c(\lambda,\lambda';m)\,,
  \end{aligned}
\end{equation}
and the relation
\begin{equation}
  \label{eq:deIf2}
  \begin{aligned}
    m^2\f{\de}{\de m^2} \Iu[\ff^2]&= \In[\ff^2] + 2m^2
    \Id[\ff^2,\ff]\,,
  \end{aligned}
\end{equation}
that follows from the last equation in Eq.~\eqref{eq:mder} and
$\de \ff / \de m^2 = -\ff^2$, one finds the equivalent set of
quantities
\begin{equation}
  \label{eq:secondorder}
  \begin{aligned}
    \coeff_{uu}&= \f{1}{m^2}\left(4m^2\Id[\ff,\ff] +
      \f{\chit-b_{N_0^2}}{m^2}+ 2m^2\f{\de}{\de m^2}\f{n_0}{m^2}+
      2m^2  \coeff_{\tilde{u}^2}\right)\,,\\
    \f{1}{2}\left( \coeff_{uu} - \coeff_{uw}\right) &= \f{\de}{\de
      m^2}\left(2m^2\coeff_{\tilde{u}^2}
      +  \f{\chit}{m^2}\right)\,,\\
    \f{1}{2}\left( \coeff_{uu} +2 \coeff_{uw} + \coeff_{ww}\right) &=
    8\Id[\ffh,\ffh]\,,\\
    \coeff_{u\tilde{u}^2} &= \f{1}{2m^2}\left( -\f{\de}{\de
        m^2}\f{\chit}{m^2} + \f{1}{2}\left( \coeff_{uu} -
        \coeff_{uw}\right) -2\coeff_{\tilde{u}^2}\right)
    \,,\\
    \f{1}{4}\left( \coeff_{uu}- \coeff_{ww}\right) - m^2\left(
      \coeff_{u\tilde{u}^2}+ \coeff_{w\tilde{u}^2}
    \right) &= \f{2}{m^2}\Iq[\ffh]\,,\\
    \coeff_{\tilde{u}^2\tilde{u}^2} &= \f{1}{m^6}\left(
      \f{b_{Q^4}-\chit}{12m^2} -\f{m^2}{16}\left( \coeff_{uu} -2
        \coeff_{uw} + \coeff_{ww}-8\coeff_{\tilde{u}^2} \right) +
      \f{m^4}{2} \left( \coeff_{u\tilde{u}^2}- \coeff_{w\tilde{u}^2}
      \right) \right) \,, 
  \end{aligned}
\end{equation}
where $\ffh\equiv \ff-m^2\ff^2$.  Since the coefficients of the
expansion of $\mathcal{W}$ in $u,w,\tilde{u}^2$ must be finite in the
chiral limit for $\mathrm{SU}(2)_A$ restoration,
Eqs.~\eqref{eq:const1}, \eqref{eq:const1chi}, \eqref{eq:const2_1},
\eqref{eq:const2_2}, \eqref{eq:const2_3}, and \eqref{eq:const2_4}
follow. In particular, since
\begin{equation}
  \label{eq:derequiv}
  \begin{aligned}
    \f{\de \coeff_{\tilde{u}^2}}{\de m^2} &= \f{\de}{\de m^2}
    \f{\de}{\de \tilde{u}^2}\wwh(m^2+u,w,\tilde{u}^2)|_0 = \f{\de}{\de
      u} \f{\de}{\de \tilde{u}^2}\wwh(m^2+u,w,\tilde{u}^2)|_0 =
    \coeff_{u\tilde{u}^2} \,,
  \end{aligned}
\end{equation}
the second and fourth line in Eq.~\eqref{eq:secondorder} are
equivalent. However, finiteness of
$\f{\de \coeff_{\tilde{u}^2}}{\de m^2}$,
\begin{equation}
  \label{eq:derequiv2}
  \f{\de \coeff_{\tilde{u}^2}}{\de m^2}
  = \f{\de}{\de m^2}
  \left(\f{n_0-\chit}{2m^4} + \Iu[\ff^2]\right)\,,
\end{equation}
provides a further constraint, 
\begin{equation}
  \label{eq:derequiv3}
  \begin{aligned}
    \f{n_0-\chit}{m^2} + 2m^2\Iu[\ff^2] &= d_1m^2 + d_2 m^4 + o(m^4) =
    2\coeff_{\tilde{u}^2}|_{m=0}\, m^2 + 2\coeff_{u\tilde{u}^2}|_{m=0}\,
    m^4+ o(m^4)\,.
  \end{aligned}
\end{equation}

\subsection{Zero modes}
\label{sec:zm}

If $\la N_+ N_-\ra= 0$, then $\la N_0^2\ra = \la Q^2\ra$, and so if
$\chit<\infty$
\begin{equation}
  \label{eq:zerom1}
  \begin{aligned}
    \f{\la N_0\ra^2}{\lvol }&\le \f{\la N_0^2\ra}{\lvol} = \f{\la
      Q^2\ra}{\lvol} \,,&&& n_0 &=\lim_{\lvol\to \infty} \f{ \la
      N_0\ra}{\lvol} \le \lim_{\lvol\to \infty} \left(\f{ \chit}{\lvol
      }\right)^{\f{1}{2}} =0\,.
  \end{aligned}
\end{equation}
Similarly,
\begin{equation}
  \label{eq:zerom2}
  \begin{aligned}
    b_{N_0^2}-\chit &= \lim_{\lvol\to \infty} \f{ \la N_0^2\ra-\la
      N_0\ra^2-\la Q^2\ra}{\lvol} = - \lim_{\lvol\to \infty} \f{\la
      N_0\ra^2}{\lvol}\,.
  \end{aligned}
\end{equation}
Conclusions are unchanged if the condition is relaxed to
$\la N_+ N_-\ra=o(\lvol)$.

\section{Spectral density}
\label{sec:app_specdens}

\subsection{Spectral density with power-series expansion around
  $\lambda=0$}
\label{sec:sp_smooth}

Assume that $\rho(\lambda;0)$ exists and that, at least for
sufficiently small values of $m$, $\rho(\lambda;m)$ admits a
convergent power expansion in $\lambda$ within a mass-independent
finite radius $\epsilon$ around $\lambda=0$,
\begin{equation}
  \label{eq:app_sppow1}
\rho(\lambda;m) = \sum_{n=0}^\infty\rho_n(m)\lambda^n\,,  
\end{equation}
with $\rho_n(m)$ finite in the chiral limit.  One has (for
$\delta<\epsilon$)
\begin{equation}
  \label{eq:app_sppow2}
  \begin{aligned}
    \lim_{m\to 0} \coeff_u & = \int_\delta^2 d\lambda \,
    \f{h(\lambda)}{\lambda^2} \rho(\lambda,0) + \int_0^\delta d\lambda
    \, \left(-\f{1}{4}\left(\rho_0(0)+\rho_1(0)\lambda\right) +
      \hh(\lambda) \sum_{n=0}^\infty\rho_{n+2}(0)\lambda^{n} \right) +
    \lim_{m\to 0} \int_0^\delta d\lambda \, \f{\rho_0(m) +
      \rho_1(m)\lambda}{\lambda^2 +m^2\hh(\lambda)}\,,
\end{aligned}
\end{equation}
and only the last term is potentially divergent. Explicitly,
\begin{equation}
  \label{eq:app_sppow3}
  \begin{aligned}
    \int_0^\delta d\lambda \, \f{1}{\lambda^2
      +m^2\left(1-\f{\lambda^2}{4}\right)}&= \f{\pi}{2 |m|} + O(1) \,,
    &&& \int_0^\delta d\lambda \, \f{\lambda}{\lambda^2
      +m^2\left(1-\f{\lambda^2}{4}\right)} &= \ln\f{1}{|m|} + O(1) \,.
  \end{aligned}
\end{equation}
Finiteness of $\coeff_u$ in the chiral limit requires then
\begin{equation}
  \label{eq:app_sppow4}
  \rho_0(m)\f{\pi}{2} + \rho_1(m) |m|\ln\f{1}{|m|} = O(m)\,.  
\end{equation}
If $\rho_1(m)=O(1/\ln\f{1}{|m|})$, then $\rho_0(m)=O(m)$ and
finiteness is obvious. If not, then $\rho_1(m)\ln\f{1}{|m|}\to\infty$
as $m\to 0$ and $\rho_1(m)|m|\ln\f{1}{|m|}$ dominates over the $O(m)$
term on the right-hand side, so that (for small enough $m$)
$\rho_1(m)\le 0$, since $\rho_0(m) = \rho(0;m)\ge 0$. One has for the
spectral density
\begin{equation}
  \label{eq:app_sppow6}
  \rho(\lambda;m) = -  \f{2}{\pi}\rho_1(m) |m|\ln\f{1}{|m|} +   O(m)  +
  \rho_1(m)\lambda + O(\lambda^2)\,.
\end{equation}
Take now
$\lambda= \lambda_\epsilon(m)\equiv |m|\ln\f{1}{|m|}
\f{2}{\pi}(1+\epsilon)$ with $\epsilon >0$. Since
$\lambda_\epsilon(m)=o(m^0)$, and
\begin{equation}
  \label{eq:app_sppow7}
  \f{\lambda_\epsilon(m)^2}{\rho_1(m) |m|\ln\f{1}{|m|}} = 
  O\left(\f{m^2\ln^2\f{1}{|m|}   }{\rho_1(m) |m|\ln\f{1}{|m|}}\right)
  = o(|m|\ln^2\tf{1}{|m|})= o(m^0)\,,
\end{equation}
one has 
\begin{equation}
  \label{eq:app_sppow8}
  \rho\left(\lambda_\epsilon(m);m\right) =
  -  \f{2}{\pi}\rho_1(m) |m|\ln\f{1}{|m|} \left(
    -\epsilon + 
    o(m^0)\right)\,,
\end{equation}
but then
\begin{equation}
  \label{eq:app_sppow9}
  0\le \lim_{m\to 0} \f{\rho\left(\lambda_\epsilon(m);m\right)
  }{-\f{2}{\pi}\rho_1(m) |m|\ln \f{1}{|m|}} =  
  -\epsilon < 0\,, 
\end{equation}
which is a contradiction.  One must then have $\rho_0(m)=O(m)$ and
$\rho_1(m)=O(1/\ln\f{1}{|m|})$. Plugged into the expression for
$\Delta$, this result gives
\begin{equation}
  \label{eq:app_sppow10}
  \begin{aligned}
    \Delta &= \lim_{m\to 0} 2\sum_{n=0}^3
    \rho_n(m)|m|^{n-1}\int_0^{\f{\delta}{|m|}} dz\, \f{
      z^n}{\left(z^2+1 \right)^2} = \lim_{m\to 0} 2 \sum_{n=0}^1
    \rho_n(m)|m|^{n-1} \int_0^\infty dz\,
    \f{z^n}{\left(z^2+1\right)^2} = \f{\pi}{2}\lim_{m\to 0}
    \f{\rho_0(m)}{|m|} \,,
  \end{aligned}
\end{equation}
so $\Delta\neq 0$ only if $\rho_0(m)= \kappa |m| + o(m)$ with
$\kappa\neq 0$. This is not possible if restoration is nonlocal, since
in that case $\rho$ is infinitely differentiable in $m^2$ at $m=0$.

\subsection{Spectral density with power-law behavior near $\lambda=0$}
\label{sec:sp_na}

I now discuss the case when the spectral density of near-zero modes is
dominated by a (possibly singular) power-law behavior. I assume that
$\rho(\lambda;0)$ exists;
$\rho(\lambda;m) = C(m)\lambda^{\alpha(m)} + \lambda r(\lambda;m)$
with $-1< \alpha(m)<1$, at least for small but finite $m$;
$r(0;m)=O(1/\ln\f{1}{|m|})$, and $u(\lambda;m)= r(\lambda;m)-r(0;m) $
is such that $|u(\lambda;m)/\lambda|< c \lambda^{\gamma}$ for some
constant $c$ and $\gamma>-1$. This set of rather loose assumptions on
the background $\lambda r(\lambda;m)$ guarantees that this gives a
finite contribution to $\coeff_u$, avoiding the need for contrived
cancellations in order to achieve symmetry restoration.  One has for
arbitrary $\delta>0$
\begin{equation}
  \label{eq:nasp1}
  \begin{aligned}
    \lim_{m\to 0} \coeff_u &= \int_\delta^2 d\lambda\,
    \f{h(\lambda)}{\lambda^2} \rho(\lambda;0) + \lim_{m\to 0}\left[
      \int_0^\delta d\lambda \,\ff(\lambda;m)\lambda r(\lambda;m) +
      C(m) \int_0^2 d\lambda
      \,\ff(\lambda;m)\lambda^{\alpha(m)}\right]\,,
  \end{aligned}
\end{equation}
where the first term is manifestly finite. For the second term
\begin{equation}
  \label{eq:nasp2}
  \begin{aligned}
    \lim_{m\to 0} \int_0^\delta d\lambda \,\ff(\lambda;m)\lambda
    r(\lambda;m) &= \lim_{m\to 0} \left[r(0;m) \left(\ln
        \f{\delta}{|m|} + o\left(\ln \tf{1}{|m|}\right)\right) +
      \int_0^\delta d\lambda \, \lambda^2 \ff(\lambda;m)
      \f{u(\lambda;m)}{\lambda}\right] <\infty\,,
  \end{aligned}
\end{equation}
since
\begin{equation}
  \label{eq:nasp2_bis}
  \left|  \int_0^\delta d\lambda \,
    \lambda^2 \ff(\lambda;m)
    \f{u(\lambda;m)}{\lambda}\right| \le 
  \int_0^\delta d\lambda \, 
  \f{\lambda^2}{\lambda^2 + m^2}
  \left|    \f{u(\lambda;m)}{\lambda}\right| \le 
  c \int_0^\delta d\lambda \, 
  \lambda^\gamma
  = \f{c\delta^{\gamma+1}}{\gamma+1}\,,
\end{equation}
while for the third term
\begin{equation}
  \label{eq:nasp3}
  \begin{aligned}
    C_0 &\equiv \lim_{m\to 0} C(m) \int_0^2 d\lambda
    \,\ff(\lambda;m)\lambda^{\alpha(m)} = \lim_{m\to 0} C(m)
    |m|^{\alpha(m)-1} \int_0^{\f{2}{|m|}} dz\, \f{z^{\alpha(m)}}{z^2
      +1}
    \\
    &= \left\{
      \begin{aligned}
        &\lim_{m\to 0} \f{ C(m) |m|^{\alpha(m)-1} \f{\pi}{2}}{\cos
          \f{\pi}{2}\alpha(m)} \,,
        &&& &\text{if }\alpha(0)\neq 1\,,\\
        &\lim_{m\to 0} C(m) \ln\f{2}{|m|} \,, &&& &\text{if
        }\alpha(0)= 1 \,.
      \end{aligned}
    \right.
  \end{aligned}
\end{equation}
The request from chiral symmetry restoration is $|C_0|<\infty$,
implying
\begin{equation}
  \label{eq:nasp8_0}
  C(m) =\left\{  \begin{aligned}
      & \f{2C_0 }{\pi}|m|^{1-\alpha(0)}   \cos
      \left(\tf{\pi}{2}\alpha(0) \right)\bar{C}(m)\,,
      &&& &\text{if } \alpha(0)\neq \pm 1\,,\\
      & \f{C_0}{\ln \f{2}{|m|}}\bar{C}(m)\,,
      &&& &\text{if } \alpha(0)= 1\,,\\
      & C_0m^2 (1+\alpha(m))\bar{C}(m)\,,
      &&& &\text{if } \alpha(0)= -1\,,
    \end{aligned}\right.
\end{equation}
with $\bar{C}(m)=1 + o(1)$. Under the same assumptions
\begin{equation}
  \label{eq:nasp4}
  \begin{aligned}
    \Delta &= \lim_{m\to 0} 2m^2 \left[\int_\delta^2 d\lambda \,
      \f{h(\lambda)^2}{\lambda^4} \rho(\lambda;0) + \int_0^\delta
      d\lambda \,\ff(\lambda;m)^2\lambda r(\lambda;m) + C(m)
      \int_0^\delta d\lambda
      \,\ff(\lambda;m)^2\lambda^{\alpha(m)}\right] \,.
  \end{aligned}
\end{equation}
The contribution of the first term manifestly vanishes. For the second
term
\begin{equation}
  \label{eq:nasp5}
  \begin{aligned}
    \lim_{m\to 0} m^2\int_0^\delta d\lambda \,\ff(\lambda;m)^2\lambda
    r(\lambda;m) &= \lim_{m\to 0} r(0;m) \left(\f{1}{2} +
      o\left(1\right)\right) + m^2\int_0^\delta d\lambda \,
    \lambda^2\ff(\lambda;m)^2 \f{u(\lambda;m)}{\lambda} = 0\,,
  \end{aligned}
\end{equation}
since
\begin{equation}
  \label{eq:nasp5_bis}
  \begin{aligned}
    \left|\int_0^\delta d\lambda \, \lambda^2 \ff(\lambda;m)^2
      \f{u(\lambda;m)}{\lambda}\right|&\le \int_0^\delta d\lambda \,
    \f{\lambda^2 }{\left(\lambda^2 + m^2\right)^2}
    \left|\f{u(\lambda;m)}{\lambda}\right|\le c\int_0^\delta d\lambda
    \, \f{\lambda^\gamma}{\lambda^2 + m^2}\,, 
  \end{aligned}
\end{equation}
and the last integral is $O(m^0)$ if $\gamma>1$, $O(\ln |m|)$ if
$\gamma =1$, and $O(|m|^{\gamma-1})$ if $-1<\gamma < 1$. Finally,
\begin{equation}
  \label{eq:nasp6}
  \begin{aligned}
    \lim_{m\to 0} 2m^2 C(m) \int_0^\delta d\lambda
    \,\ff(\lambda;m)^2\lambda^{\alpha(m)} &= \lim_{m\to 0} 2 C(m)
    |m|^{\alpha(m)-1} \int_0^{\f{\delta}{|m|}} dz \,
    \f{z^{\alpha(m)}}{(z^2 +1)^2}
    \\
    &=\left\{
      \begin{aligned}
        &\lim_{m\to 0} \f{2 C(m) |m|^{\alpha(m)-1} \f{\pi}{2} }{\cos
          \f{\pi}{2}\alpha(m)} \f{1-\alpha(m)}{2}= C_0(1-\alpha(0))\,,
        &&& &\text{if }\alpha(0)\neq 1\,,\\
        &\lim_{m\to 0} 2 C(m) |m|^{\alpha(m)-1} \f{1}{2}= \lim_{m\to
          0} O\left( \f{e^{-(1-\alpha(m))\ln\f{1}{|m|}}}{\ln
            \f{1}{|m|}}\right) = 0\,, &&& &\text{if }\alpha(0) =1\,,
      \end{aligned}\right.
  \end{aligned}
\end{equation}
having assumed in the second case that
$1-\alpha(m)= O\left(1/\ln\f{1}{|m|}\right)$. Then
\begin{equation}
  \label{eq:nasp7}
  \Delta = C_0(1-\alpha(0))\,,
\end{equation}
which includes also the case $\Delta = 0$ if $\alpha(0)= 1$.

Notice that if $-1<\alpha(0)\le 1$, the requirement of finiteness of
$\coeff_u$ in the chiral limit, Eq.~\eqref{eq:nasp8_0}, prevents
$\rho$ from being infinitely differentiable in $m^2$ at $m=0$ if
$C_0\neq 0$. For $\alpha(0)=1$ this follows immediately from the
presence of a logarithm in $C(m)$. For $-1<\alpha(0)< 1$, if
$m^{1-\alpha(0)} \bar{C}(m) = \zeta(m) +\sum_{n=0}^\infty p_n m^{2n}
$, where $d^k \zeta/dm^k|_{m=0}=0$, $\forall k$, then surely $p_0=0$.
Let $p_{\bar{n}}\neq 0$, $\bar{n}\ge 1$ be the first nonzero
coefficient. Since $1+\alpha(0)>0$, one has
$\bar{C}(m) = m^{\alpha(0)-1}\zeta(m) + m^{2(\bar{n}-1) +
  1+\alpha(0)}p_{\bar{n}}(1+o(m^0)) = o(m^0)$, which contradicts
$\bar{C}(0) =1$.  If $\alpha(0)=-1$, instead, $m^2$-differentiability
requires $\alpha$ and $C$ to be $m^2$-differentiable, which is
compatible with $C_0\neq 0$ if $\bar{C}$ if
$m^2$-differentiable. Setting
$1+\alpha(m)=\zeta_1(m)+\sum_{n=1} a_n m^{2n} $ and
$\bar{C}(m)- 1=\zeta_2(m)+\sum_{n=1}c_n m^{2n}$, where
$d^k \zeta_{1,2}/dm^k|_{m=0}=0$, $\forall k$, one finds using
Eqs.~\eqref{eq:nasp8_0} and \eqref{eq:nasp7}
\begin{equation}
  \label{eq:nasp9}
  \begin{aligned}
    C(m)&= C_0 m^2 \left(1+\alpha(m) \right)\bar{C}(m)= \f{\Delta}{2}
    m^2\left(\zeta_1(m) + m^2\sum_{n=0}^\infty a_{n+1}m^{2n}\right)
      \left(1+\zeta_2(m)+\sum_{n=1}^\infty c_n m^{2n}\right) \\ &= \f{1}{2}\Delta
    m^2\left(\zeta_1(m) + m^2a_1 + O(m^2)\right) = O(m^4) \,.
  \end{aligned}
\end{equation}
For the density of modes in the singular peak, i.e., the integral of
the singular part of $\rho$, one has (accounting also for the modes of
$D$ with negative imaginary part)
\begin{equation}
  \label{eq:nasp9bis}
  \lim_{m\to  0}\f{2}{m^2}   \int_0^2 d\lambda \,C(m) \lambda^{\alpha(m)} = \lim_{m\to
    0} 2 C_0 
  \left(1+\alpha(m) \right)\bar{C}(m)\f{2^{1+\alpha(m)}}{1+\alpha(m)}
  = \Delta\lim_{m\to
    0}   \bar{C}(m)2^{1+\alpha(m)}= \Delta =\lim_{m\to
    0} \f{\chit}{m^2}\,.
\end{equation}
In the calculations above one needs the integrals
\begin{equation}
  \label{eq:nasp10}
  \begin{aligned}
    \mathcal{I}_\gamma(\Lambda)&\equiv \int_0^\Lambda dz\, \f{
      z^{\gamma}}{z^2+1}\,,&&& \mathcal{J}_\gamma(\Lambda)&\equiv
    \int_0^\Lambda dz\, \f{ z^{\gamma}}{\left(z^2+1\right)^2}=
    \f{1-\gamma}{2}\mathcal{I}_\gamma(\Lambda)+ d_\gamma(\Lambda)\,,
    &&& d_\gamma(\Lambda)&\equiv
    \f{1}{2}\f{\Lambda^{1+\gamma}}{1+\Lambda^2} \,,
  \end{aligned}
\end{equation}
for $-1<\gamma\le 1$ and $\Lambda > 1$.  The integral
$\mathcal{I}_\gamma(\Lambda)$ can be evaluated using the residue
theorem on a half-circular contour of radius $\Lambda$ centered at the
origin (excluding an infinitesimally small half-circle around zero
that gives no contribution to the integral), obtaining
\begin{equation}
  \label{eq:nasp11}
  \begin{aligned}
    & \f{2}{\pi}\cos\left(\f{\pi}{2}\gamma\right)
    \mathcal{I}_\gamma(\Lambda) = 1-
    \f{\mathcal{R}_\gamma(\Lambda)}{\Lambda^{1-\gamma}} \,, &&& &
    \mathcal{R}_\gamma(\Lambda) \equiv \f{1}{2\pi}\int_{-\pi}^{\pi}
    d\theta \, \f{e^{-i\theta\f{1-\gamma}{2}}}{1-
      \f{1}{\Lambda^2}e^{-i\theta}}\,.
  \end{aligned}
\end{equation}
The integral $\mathcal{R}_\gamma(\Lambda)$ has a finite limit as
$\Lambda\to \infty$ for $-1\le \gamma\le 1$. If $\alpha(0)\neq 1$,
then
$(|m|/\delta)^{1-\alpha(m)}\mathcal{R}_{\alpha(m)}(\delta/|m|)\to 0$
as $m\to 0$, and $d_{\alpha(m)}(\delta/|m|)\to 0$, for any $\delta>0$,
so
\begin{equation}
  \label{eq:nasp12}
  \begin{aligned}
    \mathcal{I}_{\alpha(m)}\left(\tf{\delta}{|m|}\right) &= \f{1 +
      o(m^0)}{\f{2}{\pi}\cos\f{\pi}{2}\alpha(m)}\,, &&&
    \mathcal{J}_{\alpha(m)}\left(\tf{\delta}{|m|}\right) &=
    \f{1-\alpha(m)}{2}\f{1 +
      o(m^0)}{\f{2}{\pi}\cos\f{\pi}{2}\alpha(m)} + o(m^0)\,,
\end{aligned}
\end{equation}
independently of $\delta$. 
If $\alpha(0)=1$, setting $1-\alpha(m)=\epsilon(m)$ one finds
\begin{equation}
  \label{eq:nasp12_bis}
  \mathcal{R}_{\alpha(m)}\left(\tf{\delta}{|m|}\right)
  = 1 +
  \epsilon(m) O\left(\epsilon(m), \tf{m^2}{\delta^2}\right)\,,
\end{equation}
so if $\epsilon  =o\left(1/\ln \f{1}{|m|}\right)$
\begin{equation}
  \label{eq:nasp12_ter}
  \f{ \mathcal{I}_{1-\epsilon(m)}\left(\tf{\delta}{|m|}\right)}{\ln
    \tf{\delta}{|m|}}
  = \f{1- e^{-\epsilon(m)\ln\tf{\delta}{|m|}}\left(1 +
      \epsilon(m)   O\left(\epsilon(m), \tf{m^2}{\delta^2}\right)\right)}{
    \epsilon(m) \ln\tf{\delta}{|m|} (1+ O(\epsilon(m)^2))}\mathop
  \to_{m\to 0} 1\,.
\end{equation}
For $\mathcal{J}_{\alpha(m)}$ one has then
$\lim_{m\to 0} \mathcal{J}_{\alpha(m)}\left(\tf{\delta}{|m|}\right)=
\mathcal{J}_{1}\left(\infty\right)=\f{1}{2}$.

\section{Two-point function}
\label{sec:app_2pfunc}

\subsection{Finite $\rho_c^{(2)}(0,0;m)$}
\label{sec:tp_fin}

Assuming
$\rho_c^{(2)}(\lambda,\lambda';m)=A(m)+B(\lambda,\lambda';m)$, with
$B$ integrable, vanishing at the origin, and obeying the bound
$|B(\lambda,\lambda';m)|\le b (\lambda^2 +
\lambda^{\prime\,2})^{\f{\beta}{2}}$ for $b>0$, $0<\beta< 1$, one has
\begin{equation}
  \label{eq:twop1}
  \begin{aligned}
    \lim_{m\to 0} m^2 \Id[\ff,\ff] &= \lim_{m\to 0} \left[A(m) \left(m
      \int_0^2d\lambda \,\ff(\lambda;m)\right)^2 + I_B(m)\right] =
    A(0)\mathcal{I}_0(\infty)^2 + \lim_{m\to 0} I_B(m)
    \,,\\
    I_B(m) & \equiv m^2\int_0^2d\lambda \int_0^2
    d\lambda'\,\ff(\lambda;m)\ff(\lambda';m) B(\lambda,\lambda';m)\,,
  \end{aligned}
\end{equation}
where $\mathcal{I}_\gamma$ is defined in Eq.~\eqref{eq:nasp10}.
Using the bound on $B$,
\begin{equation}
  \label{eq:twop2}
  \begin{aligned}
    | I_B(m)| &\le bm^2 \int_0^2d\lambda \int_0^2
    d\lambda'\,\ff(\lambda;m)\ff(\lambda';m) \left(\lambda^2 +
      \lambda^{\prime\,2}\right)^{\f{\beta}{2}} \le bm^{\beta}
    \int_0^\infty dz \int_0^\infty dz'\,\f{\left(z^2 +
        z^{\prime\,2}\right)^{\f{\beta}{2}}}{(z^2+1)(z^{\prime\,
        2}+1)} \\ &= bm^{\beta}\int_0^{\f{\pi}{2}}d\phi \int_0^\infty
    dr\,\f{r^{1+\beta}}{ \tf{1}{4}r^4\left(\sin\phi\right)^2 +r^2+1}
    \le bm^{\beta}\int_0^{\f{\pi}{2}}d\phi\left\{ 1 +
      \left(\f{2}{\sin\phi}\right)^\beta \int_0^{\infty}
      dr\,\f{r^{1-\beta}}{r^2+1} \right\} \,.
  \end{aligned}
\end{equation}
The last integral is finite, so $\lim_{m\to 0} I_B(m)=0$. Then, using
$\mathcal{I}_0(\infty) = \f{\pi}{2}$, one finds
$\lim_{m\to 0} 4m^2 \Id[\ff,\ff] = \pi^2 A(0)$. Similarly,
\begin{equation}
  \label{eq:twop4}
  \begin{aligned}
    \lim_{m\to 0} m^2 \Id[\ffh,\ffh] &= \lim_{m\to 0} \left[A(m)
      \left(m \int_0^2d\lambda \,\ffh(\lambda;m)\right)^2 +
      \hat{I}_B(m)\right] = A(0)
    \left(\mathcal{I}_0(\infty)-\mathcal{J}_0(\infty) \right)^2 +
    \lim_{m\to 0}\hat{I}_B(m)
    \,,\\
    \hat{I}_B(m) &\equiv m^2\int_0^2d\lambda \int_0^2
    d\lambda'\,\ffh(\lambda;m)\ffh(\lambda';m)
    B(\lambda,\lambda';m)\,,
\end{aligned}
\end{equation}
where $\mathcal{J}_\gamma$ is defined in Eq.~\eqref{eq:nasp10}. Since
\begin{equation}
  \label{eq:twop5}
  \begin{aligned}
    | \hat{I}_B(m)| &\le bm^2 \int_0^2d\lambda \int_0^2
    d\lambda'\,\ff(\lambda;m)\ff(\lambda';m) \left(\lambda^2 +
      \lambda^{\prime\,2}\right)^{\f{\beta}{2}}\,,
  \end{aligned}
\end{equation}
it follows from Eq.~\eqref{eq:twop2} that
$\lim_{m\to 0}\hat{I}_B(m)=0$. Using
$\mathcal{J}_0(\infty) = \f{\pi}{4}$, one finds
$\lim_{m\to 0} m^2 \Id[\ffh,\ffh] = \f{\pi^2}{16}A(0)$.

\subsection{Localized near-zero modes}
\label{sec:2p_loc}

Assume
$|\rho_c^{(2)}(\lambda,\lambda';m)| \le C
\rho(\lambda;m)\rho(\lambda';m)$ for $\lambda,\lambda'<\lambda_c$, and
$|\rho_c^{(2)}(\lambda,\lambda';m)| \le C' \rho(\lambda;m) $ for
$\lambda<\lambda_c<\lambda'$, for some mass-independent constants
$C,C'>0$; and that $|\rho_c^{(2)}(\lambda,\lambda';m)|$ is integrable
$\forall m\ge 0$ for $\lambda,\lambda'>\lambda_c$.  Assume also that
$\lim_{m\to 0}\lambda_c\neq 0$. Then
\begin{equation}
  \label{eq:singtwop1}
  \begin{aligned}
    \left|\lim_{m\to 0} m^2 \Id[\ff,\ff]\right| & \le \lim_{m\to 0}
    \left[C m^2 \left( \int_0^{\lambda_c}d\lambda\,\ff(\lambda;m)
        \rho(\lambda;m)\right)^2 + 2 C' m^2\int_0^{\lambda_c}d\lambda
      \,\ff(\lambda;m) \rho(\lambda;m) \int_{\lambda_c}^2
      d\lambda'\,\ff(\lambda';m) \right]
    \\
    &\le \lim_{m\to 0} m^2 \left(C \coeff_u^2 + 2 C' \coeff_u I_L(m)
    \right)\,,
  \end{aligned}
\end{equation}
where
\begin{equation}
  \label{eq:singtwop2}
  \begin{aligned}
    I_L(m) &\equiv \int_{\lambda_c}^2 d\lambda\,\ff(\lambda;m)\,,&&&
    \lim_{m\to 0} I_L(m) &= \lim_{m\to 0} \int_{\lambda_c}^2 d\lambda
    \f{\hh(\lambda)}{\lambda^2} =\lim_{m\to 0} \f{1}{\lambda_c}
    \left(1-\f{\lambda_c}{2}\right)^2 <\infty\,.
  \end{aligned}
\end{equation}
It follows that $\lim_{m\to 0} m^2 \Id[\ff,\ff]=0$ if
$\lim_{m\to 0} \coeff_u<\infty$. Since $0\le \ffh\le \ff$ for
$\lambda\in [0,2]$, a similar calculation shows that
\begin{equation}
  \label{eq:singtwop4}
  \begin{aligned}
    \left|\lim_{m\to 0} \Id[\ffh,\ffh]\right| &\le \lim_{m\to 0}
    \left[C \coeff_u^2 + 2 C' \coeff_u I_L(m) +
      \f{1}{\lambda_c^4}\int_{\lambda_c}^2d\lambda \int_{\lambda_c}^2
      d\lambda' \, |\rho_c^{(2)}(\lambda,\lambda';m)| \right]
    <\infty\,.
  \end{aligned}
\end{equation}

\end{widetext}